\documentclass[12pt]{iopart}
\usepackage[totalwidth=15.7cm, totalheight=24.0cm]{geometry}
\usepackage{graphicx,times}
\newcommand{\sauron}{\texttt{SAURON}}
\begin{document}

\title[Revisiting the $(V/\sigma)-\varepsilon$ diagram]{Revisiting the $(V/\sigma)-\varepsilon$ anisotropy diagram of early-type galaxies using integral-field kinematics}

\author{M.~Cappellari,$^1$ R.~Bacon,$^2$ M.~Bureau,$^3$ R.~L.~Davies,$^3$ P.~T.~de~Zeeuw,$^1$ E.~Emsellem,$^2$ J.~Falc\'on-Barroso,$^1$ D.~Krajnovi\'c,$^3$ H.~Kuntschner,$^4$ R.~M.~McDermid,$^1$ R.~F.~Peletier,$^5$ M.~Sarzi,$^3$ R.~C.~E.~van~den~Bosch$^1$ and G.~van~de~Ven$^1$}

\address{$^1$ Leiden Observatory, Postbus 9513, 2300 RA Leiden, The Netherlands}
\address{$^2$ Centre de Recherche Astronomique de Lyon, Lyon, France}
\address{$^3$ University of Oxford, Oxford, UK}
\address{$^4$ Space Telescope European Coordinating Facility, Garching, Germany}
\address{$^5$ Kapteyn Astronomical Institute, Groningen, The Netherlands}

\ead{cappellari@strw.leidenuniv.nl}

\begin{abstract}
We use integral-field observation of the stellar kinematics obtained with \sauron\ in combination with Schwarzschild dynamical models to revisit our understanding of the classic $(V/\sigma)-\varepsilon$ anisotropy diagram of early-type galaxies.
\end{abstract}

\vspace{-0.1cm}
\submitto{Nearly Normal Galaxies in a $\Lambda$CDM Universe. A conference celebrating the 60th birthdays of George Blumenthal, Sandra Faber and Joel Primack. Santa Cruz, CA, 2005}

\section{Surprising Anisotropy Results}

We analyze the orbital distribution of a sample of 25 elliptical (E) and lenticular (S0) galaxies using the axisymmetric three-integral Schwarzschild dynamical models presented in Cappellari et al.\ (2005). The models are constrained by \sauron\ (Bacon et al.\ 2001) integral-field kinematics (Emsellem et al.\ 2004) within about one effective (half-light) radius $R_{\rm e}$, and by {\em Hubble Space Telescope} and MDM ground-based photometry. The modeled sample is constituted by half of the early-type galaxies of the \sauron\ representative sample (de Zeeuw et al. 2002), and also includes the dwarf elliptical M32 to explore the low-luminosity range. This is the first time such a large and homogeneous analysis of the orbital anisotropy can be performed using integral-field stellar kinematics and relatively general models.

We use the \sauron\ kinematics to quantitatively define a new classification of early-type galaxies (Emsellem et al.\ in prep.; see McDermid et al.\ 2005), with and without a significant amount of angular momentum per unit stellar mass, in the spirit of the classification by Kormendy \& Bender (1996). We refer to these two types of object as ``fast-rotators'' and ``slow-rotators'' respectively (see Fig.~1 for an illustration).

\begin{figure}
\centering
\includegraphics[width=0.7\textwidth]{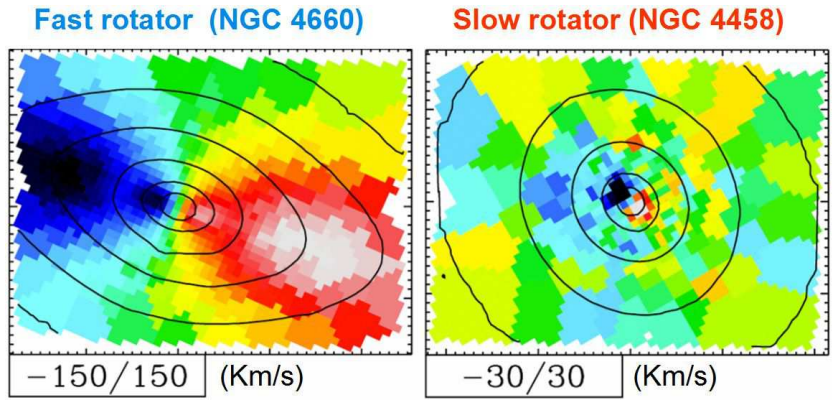}
\caption{Mean stellar velocity field ($30\times40$ arcsec) observed with \sauron\ (from Emsellem et al.\ 2004) for two prototypical galaxies in the fast-rotators and slow-rotators classes respectively.}
\end{figure}

From the analysis of the dynamical models we find that, in the central regions of the galaxies which we sample with our kinematics:
\begin{enumerate}
\item The slow-rotators are more common among the brightest systems, they are generally classified as E from photometry alone, they can display misalignement between the photometric and kinematical axes, indicating they cannot all be axisymmetric, they tend to appear nearly round and to have an almost isotropic velocity dispersion tensor. Binney's (1978) anisotropy parameter $\delta\sim0$ (left panel of Fig.~2);

\item The fast-rotators are generally fainter, can appear flattened and do not show significant kinematical misalignment, which indicates they are all nearly axisymmetric, they are classified either E or S0, and span a large range of anisotropies (up to $\delta\sim0.6$). We find a good correlation $\beta-\varepsilon_{\rm intr}$ between the anisotropy in the meridional plane and the galaxy intrinsic flattening (right panel of Fig.~2).
\end{enumerate}
{\em These results are the opposite of the classic interpretation of the $(V/\sigma)-\varepsilon$ diagram, where the bright non-rotating galaxies are anisotropic and the faint fast-rotating ones are isotropic.}

\begin{figure}
\centering
\includegraphics[height=5cm]{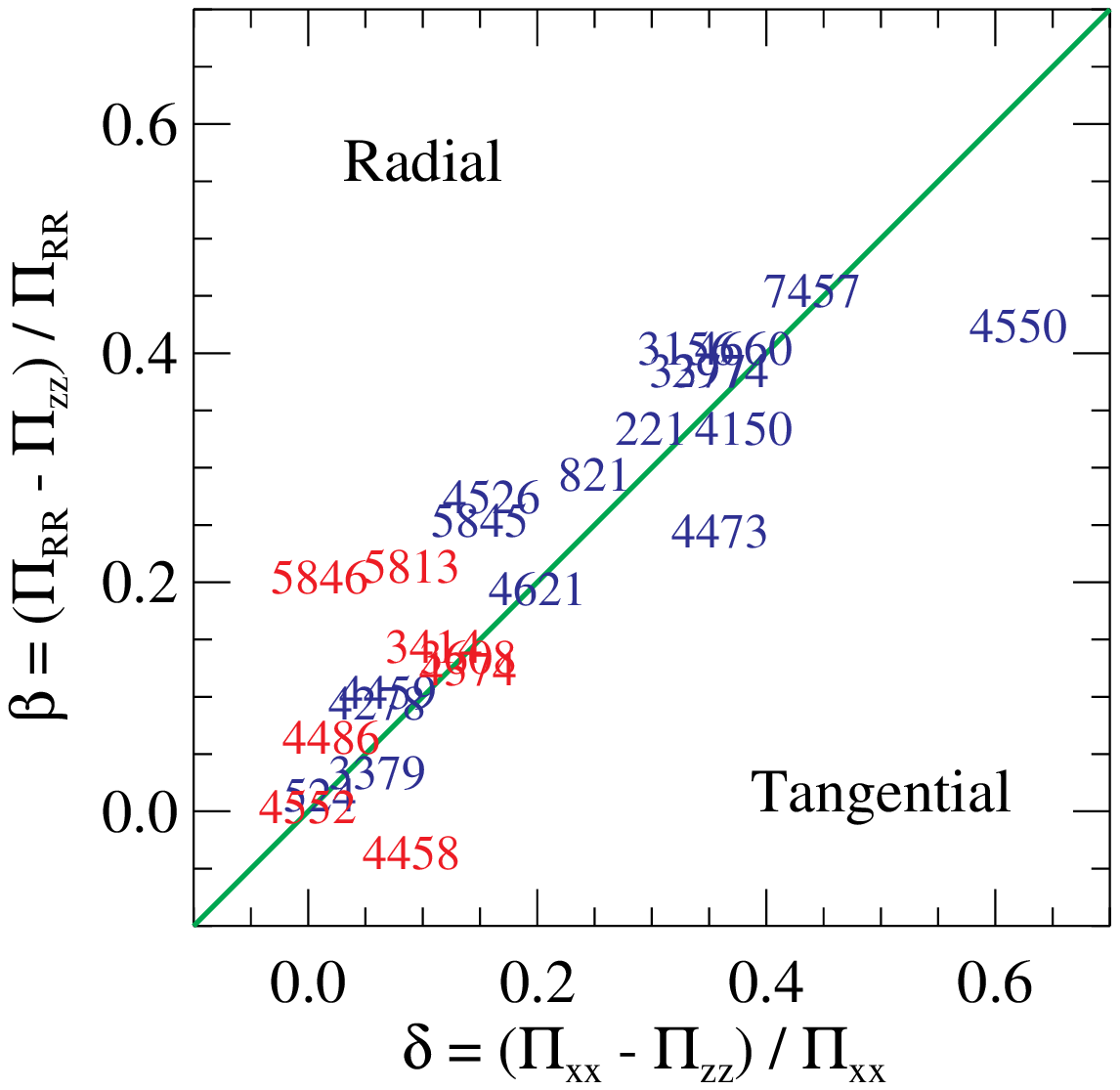}
\includegraphics[clip,trim=55 0 0 0,height=5cm]{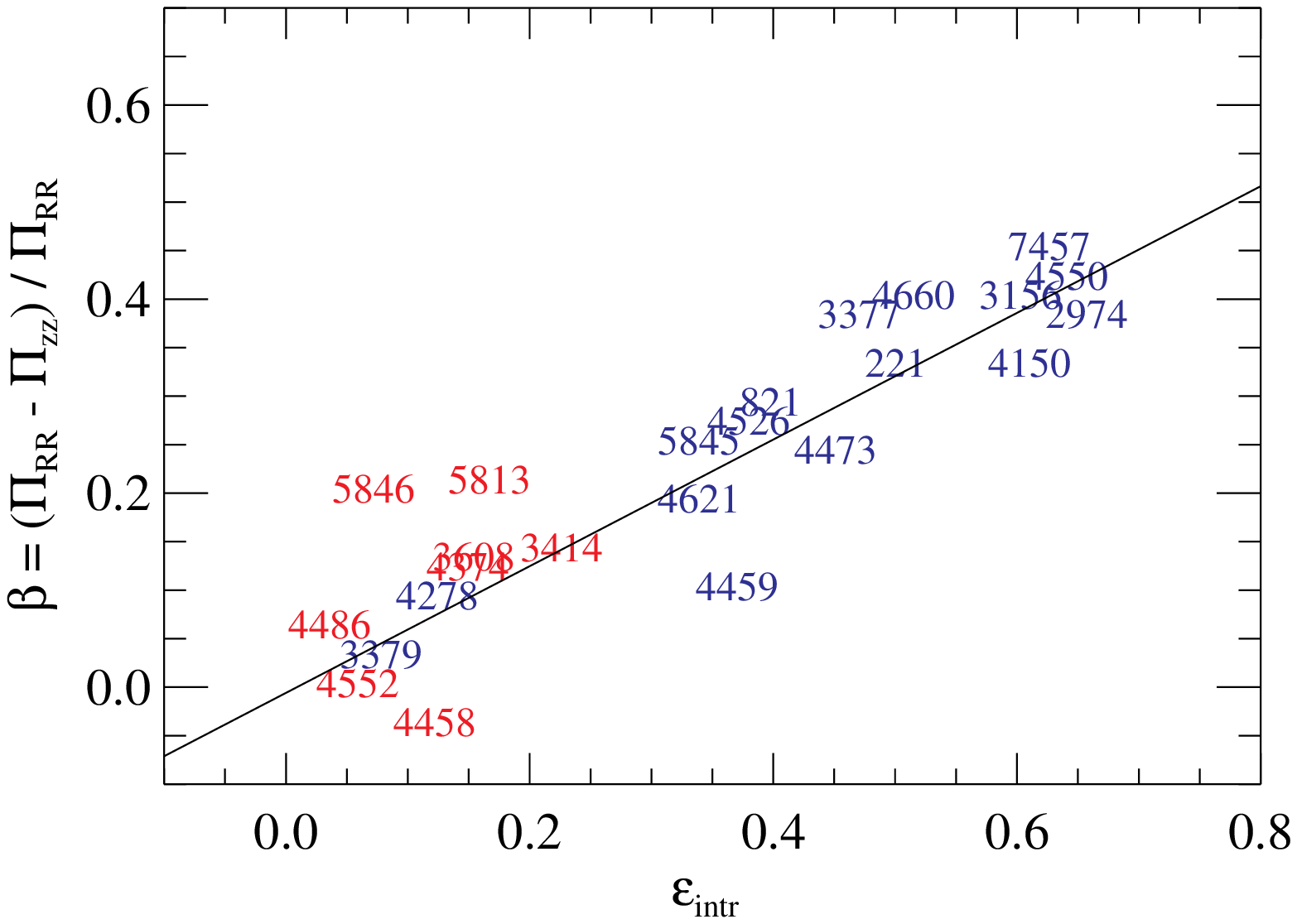}
\caption{{\em Left panel:} the mass-weighted anisotropy parameter $\beta\equiv1-\sigma_z^2/\sigma_R^2$ versus the anisotropy $\delta\equiv1-2\sigma_z^2/(\sigma_R^2+\sigma_\phi^2)$ measured from the orbital distribution of the 25 dynamical models, within the volume sampled by the observed kinematics. All objects, with the exception of the dwarf elliptical M32 (NGC~221), are early-type galaxies from the \sauron\ representative sample, and are also shown in the diagram of Fig.~3. The green line indicate galaxies with $\sigma_R=\sigma_\phi$. Most of the galaxies fall close to the green line, indicating that the anisotropy is characterized by a flattening of the velocity ellipsoid in the meridional plane, while the velocity ellipsoid tends to be circular in a plane parallel to the equatorial plane. The red and blue labels refer to the slow-rotators and fast-rotators respectively (see Fig.~1 for an illustration).
{\em Right panel:} correlation between the intrinsic flattening $\varepsilon_{\rm intr}$ and the anisotropy parameter $\beta$. The flatter galaxies tend to be more anisotropic, as one may expect if the increasing flattening is due to a disk heating process.}
\end{figure}

\section{Understanding the $(V/\sigma)-\varepsilon$ Diagram}

We measure the distribution of all the 48 E/S0 galaxies of the \sauron\ representative sample, in the classic anisotropy diagram (Illingworth 1977; Binney 1978; Davies et al.\ 1983), which relates the ratio of the ordered and random motion in a galaxy ($V/\sigma$), to its observed flattening $\varepsilon$. For the first time we can construct the $(V/\sigma)-\varepsilon$ diagram in a rigorous manner (Binney 2005) using luminosity-weighted quantities from our integral-field stellar kinematics.

\begin{figure}
\centering
\includegraphics[width=0.6\columnwidth]{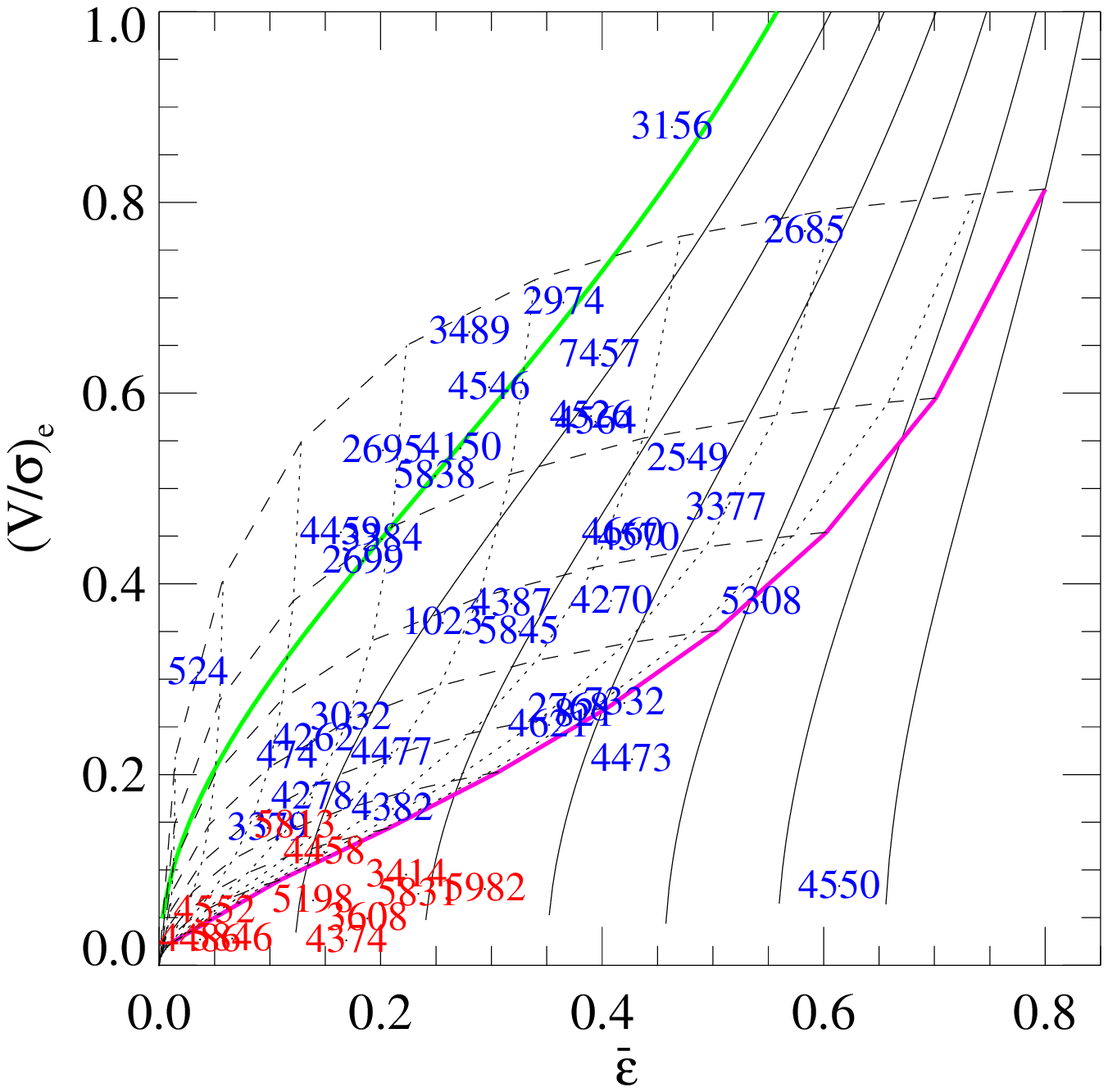}
\caption{$(V/\sigma)-\varepsilon$ diagram for all the 48 E/S0 galaxies in the \sauron\ representative sample (half of these objects are in common with Fig.~2). The kinematics were measured in a robust way from our luminosity-weighted integral-field kinematics (inside one $R_{\rm e}$) as  $(V/\sigma)_e^2\equiv\langle V^2 \rangle / \langle \sigma^2 \rangle$, following Binney (2005). The ellipticity $\bar{\varepsilon}$ is luminosity-weighted within the same region adopted for the kinematics.
The red and blue labels refer to the slow-rotators and fast-rotators respectively (see Fig.~1). The green line shows the location of oblate isotropic ($\delta=0$) edge-on galaxies, while subsequent black solid lines indicate increasing anisotropy values $\delta=0.1,0.2,\ldots,0.6$. The magenta line corresponds to a linear relation between anisotropy $\delta$ and ellipticity $\varepsilon$ for edge-on galaxies. The dotted lines show the location of galaxies, originally on the magenta line, when the inclination is decreased $i=80^\circ,70^\circ,\ldots,10^\circ$. The dashed lines correspond to constant intrinsic ellipticity in the range $\varepsilon_{\rm intr}=0.8,0.7,\ldots,0.1$. While the fast-rotators can be described as a family of oblate anisotropic galaxies, seen at different inclinations, the slow-rotators are inconsistent with this distribution. The presence of kinematical misalignments indicates that the slow-rotators are (weakly) triaxial systems. The two galaxies NGC~4473 and NGC~4550 fall below the magenta line due to the strong tangential anisotropy (left panel of Fig.~2) caused by major counterrotating stellar disks (Rix et al.\ 1992; Cappellari \& McDermid 2005).}
\end{figure}

The distribution of the galaxies in the \sauron\ $(V/\sigma)-\varepsilon$ diagram (Fig.~3) is {\em consistent} with the findings from the modeling above, when inclination effects are taken into account. The diagram appears to be populated by the two groups of (i) weakly triaxial and nearly isotropic slow-rotators (ii) nearly oblate and anisotropic fast-rotators.

The significant anisotropy of the fast-rotators agrees with the recent results of numerical simulations (Burkert \& Naab 2005) suggesting a formation process from gas rich disk galaxies, followed by disk heating due to minor mergers or secular evolution. The slow-rotators have long been considered the result of equal mass collisionless mergers (Barnes 1988; Bender, Burstein \& Faber 1992) but from our new data they appear significantly rounder and closer to isotropic than current numerical simulations predict. We are currently investigating the reasons of the differences between our new $(V/\sigma)-\varepsilon$ and the ones previously derived from long-slit spectroscopy, including selection effects.

We speculate that the two classes of fast and slow-rotating galaxies may be the relics of the different formation paths followed by early-type galaxies when loosing their gas content and moving from the ``blue cloud'' to the ``red sequence'', in the ``quenching'' scenario of Faber et al.\ (2005). In that picture the fast-rotators are naturally associated to the spiral galaxies whose disks were quenched by ram-pressure stripping or other processes not involving significant mergers. Subsequent disk heating associated to minor mergers or secular evolution can explain the correlation between the thickening of the disk and the vertical velocity dispersion (Fig.~2). The slow-rotators can be generated by a sequence of `dry' mergers (i.~e.\ involving little amount of gas) along the red sequence, or due to major violent gas rich mergers in which the gas component was rapidly expelled by a starburst or a central AGN.

\References

\item[] Bacon R., et al.\ 2001, MNRAS, 326, 23
\item[] Barnes J.~E., 1988, ApJ, 331, 699
\item[] Bender R., Burstein D., \& Faber S.~M., 1992, ApJ, 399, 462
\item[] Binney J., 1978, MNRAS, 183, 779
\item[] Binney J., 2005, MNRAS, submitted (astro-ph/0504387)
\item[] Burkert A., \& Naab T., 2005, MNRAS, submitted (astro-ph/0504595)
\item[] Cappellari M., \& McDermid R.~M., 2005, Class. Quantum Grav., 22, 347
\item[] Cappellari M., et al.\ 2005, MNRAS, in press (astro-ph/0505042)
\item[] Davies R.~L., Efstathiou G.~P., Fall S.~M., Illingworth G.~D., Schechter P.~L., 1983, ApJ, 266, 41
\item[] de Zeeuw P.~T., et al.\ 2002, MNRAS, 329, 513
\item[] Emsellem E., et al.\ 2004, MNRAS, 352, 721
\item[] Faber S.~M., et al.\ 2005, ApJ, submitted, (astro-ph/0506044)
\item[] Illingworth G., 1977, ApJ, 218, L43
\item[] Kormendy J., \& Bender R., 1996, ApJ, 464, L119
\item[] McDermid R.~M., et al.\ 2005, New Astr.\ Rev.\, in press (astro-ph/0508631)
\item[] Rix H.-W., Franx M., Fisher D., Illingworth G., 1992, ApJL, 400, 5

\endrefs

\end{document}